%% file: CPPRS.tex
\documentclass[12pt,fleqn]{article}
\pdfoutput=1
\usepackage[DIV12]{typearea}
\usepackage{amsmath,amsfonts,amssymb}
\usepackage{graphicx}
\usepackage{cite}
\usepackage{mathrsfs}
\usepackage{bbm}
\usepackage{pifont}
\usepackage{units}
\usepackage[small]{caption} 
\usepackage[small,loose]{subfigure}  
\usepackage{color}
\usepackage{nomencl}
\usepackage{psfrag}
\usepackage{longtable}

\newcommand{\eVdist}{\kern-0.06em}

\DeclareMathOperator{\tr}{tr}
\DeclareMathOperator{\diag}{diag}

\newcommand{\I}{\mathrm{i}}

\newcommand{\E}[1]{\ensuremath{\mathrm{E}_{#1}}} 

\newcommand{\SO}[1]{\ensuremath{\mathrm{SO}(#1)}}
\newcommand{\SU}[1]{\ensuremath{\mathrm{SU}(#1)}}

\newcommand{\U}[1]{\ensuremath{\mathrm{U}(#1)}}
\newcommand{\Z}[1]{\ensuremath{\mathbbm{Z}_{#1}}} 

\newcommand{\x}[0]{\ensuremath{\times}}
\newcommand{\p}[0]{\ensuremath{\cdot}}
\newcommand{\maN}{\ensuremath{\mathcal{N}}}
\newcommand{\maT}{\ensuremath{\mathbb{T}}}
\newcommand{\maR}{\ensuremath{\mathbb{R}}}
\newcommand{\maG}{\ensuremath{\mathcal{G}}}
\newcommand{\bs}[1]{\ensuremath{\boldsymbol{#1}}}
\newcommand{\rep}[1]{\ensuremath{\boldsymbol{#1}}}
\newcommand{\crep}[1]{\ensuremath{\overline{\boldsymbol{#1}}}}

\newcommand{\vev}[1]{\langle #1\rangle}

\allowdisplaybreaks[1]

\def\MyTitle{$\Delta(54)$ flavor phenomenology and strings}
\title{\MyTitle}

\begin{document}

\begin{titlepage}

\vspace*{1.0cm}

\begin{center}
{\Large\bf\MyTitle
}

\vspace{1cm}

\textbf{
Brenda Carballo--P\'erez$^{\ast\dagger}$\footnote[1]{Email: \texttt{brenda.carballo@hebaideas.com}}{},
Eduardo Peinado$^{\ast}$\footnote[2]{Email: \texttt{epeinado@fisica.unam.mx}}{},
Sa\'ul~Ramos--S\'anchez$^{\ast}$\footnote[3]{Email: \texttt{ramos@fisica.unam.mx}}{}
}
\\[5mm]
\textit{\small
$^{\ast}$Instituto de F\'isica, Universidad Nacional Aut\'onoma de M\'exico\\
Apartado Postal 20-364, Ciudad de M\'exico 01000, M\'exico\\
\vspace{0.5cm}
$^{\dagger}$HEBA Ideas S.A. de C.V.\\
Calculistas 37, Cd. Mx. 09400 M\'exico
}
\end{center}

\vspace{2cm}

\begin{abstract}
$\Delta(54)$ can serve as a flavor symmetry in particle physics, but remains almost
unexplored. We show that in a classification of semi--realistic \Z3\x\Z3 
heterotic string orbifolds, $\Delta(54)$ turns out to be the most natural flavor symmetry,
providing additional motivation for its study. We revisit its phenomenological 
potential from a low--energy perspective and subject to the constraints of string
models. We find a model with $\Delta(54)$ arising from heterotic orbifolds that leads to the 
Gatto-Sartori-Tonin relation for quarks and charged--leptons. Additionally, in the neutrino sector, 
it leads to a normal hierarchy for neutrino masses and a correlation between the 
reactor and the atmospheric mixing angles, the latter taking values in the 
second octant and being compatible at three sigmas with experimental data.
\end{abstract}

\end{titlepage}

\section{Introduction}

The standard model (SM) exhibits  features, such as the family repetition and the structure of 
mixing matrices for quarks and leptons, that suggest an underlying structure. 
Non--Abelian discrete flavor symmetries appear in many bottom--up models as a promising explanation 
for these observations~\cite{Ishimori:2010au,Hirsch:2012ym,King:2013eh}.

A large set of Abelian and non--Abelian discrete symmetries has been successfully investigated 
in this context~\cite{Aranda:1999kc,Aranda:2000tm,Aranda:2007dp,Babu:2004tn,Babu:2011mv,Babu:2002dz,Hirsch:2003dr,Ma:2001dn,Altarelli:2005yp,Altarelli:2010gt,deMedeirosVarzielas:2005qg,Grimus:2004rj,Grimus:2004hf,Morisi:2007ft,Bazzocchi:2008ej,Hirsch:2010ru,Campos:2014zaa,Hernandez:2013hea,King:2013eh,deMedeirosVarzielas:2006fc,King:2006np,Hernandez:2012ra,Ding:2009iy,Ding:2012wh}.
Particularly, the groups \Z3~\cite{Dev:2011jc,Peinado:2012tp,Sierra:2014kua,Aranda:2014lna}, 
$S_3$~\cite{Pakvasa:1977in,Frere:1978ds,Ma:1991eg,Fukugita:1998vn,Kubo:2003iw,Caravaglios:2005gw,Mondragon:2007af,Das:2015sca}  
and $\Delta(27)$~\cite{deMedeirosVarzielas:2006fc,Luhn:2007uq,Morisi:2012hu,Ferreira:2012ri,Holthausen:2012dk,Aranda:2013gga,Varzielas:2015aua,Vien:2016tmh}
have shed some light on the structure of the quark and neutrino sectors, providing in some cases an explanation of proton stability 
and dark matter~\cite{Morisi:2012hu,Chen:2013dpa,Hirsch:2010ru,Boucenna:2011tj,Lamprea:2016egz} or 
an explanation of the {\it Dirac--ness} of neutrinos~\cite{Aranda:2013gga}.
These symmetries have in common that they are subgroups of $\Delta(54)$, which
however has been explored only aiming at a tri--bimaximal neutrino--mixing structure or 
similar~\cite{Ishimori:2008uc,Ishimori:2009ew,Escobar:2011mq,Boucenna:2012qb}.
Since $\theta_{13}$ is now known to be non--zero, the potential of 
$\Delta(54)$ as a flavor symmetry must be revisited. To pave the way to a 
vast revision on this subject is one of the goals of this work.

On the other hand, despite their success, the origin of flavor symmetries remains
unexplained in bottom--up model building. Fortunately, non--Abelian flavor symmetries
emerge naturally in different compactification schemes
of string theory~\cite{Kobayashi:2004ud,Kobayashi:2006wq,Ko:2007dz,Nilles:2012cy,Beye:2014nxa,Beye:2015wka,Abe:2016eyh} 
that enjoy the properties of the SM or its supersymmetric extension(s), yielding a promising
ultraviolet completion of flavor phenomenology.

Toroidal heterotic orbifolds~\cite{Dixon:1985jw,Dixon:1986jc} (see e.g.~\cite{Bailin:1999nk} for a comprehensive introduction) 
lead to models which reproduce the gauge group and matter spectrum of the SM~\cite{Blaszczyk:2014qoa}, 
its minimal supersymmetric extension~\cite{Kobayashi:2004ya,Buchmuller:2005jr,Kim:2007mt,Blaszczyk:2009in} 
and other non--minimal extensions~\cite{Lebedev:2009ag}, as well as many other observed and/or desirable properties
of particle physics~\cite{Buchmuller:2007zd,Kappl:2008ie,Choi:2009jt,Brummer:2010fr,Krippendorf:2012ir,Badziak:2012yg,Kim:2015mpa}.
As we discuss in section~\ref{sec:symmetries}, following previous findings of~\cite{Kobayashi:2006wq,Nilles:2012cy}, 
a $\Delta(54)$ flavor symmetry can emerge in these constructions as a result of dividing a $\mathbb T^2$ torus by 
\Z3 in the compact dimensions. A paramount difference between the flavor theory emerging in
this context and one arbitrarily proposed is that all properties, including the flavor 
representations and number of fields, are dictated by the string compactification itself, resulting
in interesting phenomenological consequences that we aim at studying in this paper.

Due to their geometrical structure, \Z3 or \Z3\x\Z2 heterotic orbifolds could 
in principle yield a $\Delta(54)$ flavor symmetry, 
but it is known that no promising model where this symmetry remains unbroken arises 
in those cases~\cite{Lebedev:2006kn,Lebedev:2007hv,Lebedev:2008un}. Therefore, the simplest
complete string scenarios with SM--like physics and this flavor symmetry are \Z3\x\Z3 heterotic orbifolds. 

In this paper, we explore the phenomenological viability of the $\Delta(54)$ flavor
symmetry from a top--down and a bottom--up perspective. After explaining in section~\ref{sec:symmetries}
how flavor symmetries relate to geometry in heterotic string compactifications, in section~\ref{sec:class}
we perform a search of semi--realistic \Z3\x\Z3 heterotic orbifold models, which turn out
to display $\Delta(54)$ as a flavor symmetry more naturally than other possibilities.
In section~\ref{sec:stringyDelta54} we inspect the flavor symmetries and spectrum 
properties of one string sample model. Inspired by the features of the string models,
in section~\ref{sec:Delta54pheno} we propose a model that reproduces at some level
known flavor observations and provides predictions for the neutrino sector.
In section~\ref{sec:conclusions} we provide our concluding remarks.


\section{Origin of flavor symmetries in heterotic orbifolds}
\label{sec:symmetries}

We follow here the discussion of~\cite{Kobayashi:2006wq,Nilles:2012cy}, stressing some important aspects for our work.
 
In higher dimensional models, such as the string theories, flavor symmetries result  from
the geometrical symmetries (and other properties) of the extra dimensions (see e.g.~\cite{Altarelli:2006kg}
for a field--theoretical proposal). Since in those models
the extra dimensions must be compactified in order to justify that we only perceive four dimensions,
the compact space adopts geometrical structures which are endowed with symmetries that 
are passed down, as flavor symmetries, to the fields arising in those constructions. 

Among all possibilities, orbifolds are perhaps the simplest compactifications.
A $d$--dimensional orbifold is defined as the quotient of $\mathbb R^d$ divided by a 
discrete group. The resulting space is a compact solid, exhibiting typically 
some curvature singularities (fixed points of the orbifold), 
at which matter states may be localized. In the absence of local effects
at the singularities, the states attached to all singularities are
indistinguishable. The transformations (permutations, reflections, etc.) of those 
identical states that leave the matter distribution invariant build a (non-Abelian) 
symmetry of the compactified theory. Note that such transformations are equivalent to 
field relabelings.

As a first example, let us suppose that an orbifold yields a compact space endowed with
two singularities at which two matter generations are chosen to be localized. Since these localized
matter generations are indistinguishable, i.e. have identical quantum numbers, 
excepting of course for their localization properties, a permutation or relabeling of
the generations does not alter the system. That is, the system is invariant under 
an $S_2$ permutation symmetry, leading to an effective model with
two generations related to each other under the non--trivial (flavor)
transformation of that group.

In string theory, the simplest and yet quite promising compactifications of this kind
are toroidal heterotic orbifolds~\cite{Dixon:1985jw,Dixon:1986jc}. 
They are achieved by letting first the six extra
dimensions of a 10D heterotic string be compact by imposing the quotient
$\maR^6/\Lambda_G$, where $\Lambda_G$ can be chosen as a 6D root lattice of a 
Lie group $G$. The resulting 6D torus $\maT^6=\maR^6/\Lambda_G$ is then divided by
a discrete group of its isometries $P$, yielding the orbifold $\mathbb O = \maT^6/P$.
$\mathbb O$ is Abelian when $P$ is Abelian. For simplicity, 
we shall focus here only on Abelian orbifolds.

Not any arbitrary choice of $\mathbb T^6$ and $P$ is admissible. Requiring unbroken 
supersymmetry in the effective 4D field theory as well as considering topological 
equivalences between compactifications with different geometries
reduce greatly the number of allowed heterotic orbifolds. In fact,
all possible 6D orbifolds of this type have been exhaustively 
classified~\cite{Fischer:2012qj}, resulting in a small number of Abelian orbifolds 
and thus a small number of possible geometrical symmetries to be considered.

In contrast to a bottom--up approach, where matter fields are arbitrarily 
localized at the singularities or let free in the bulk, in heterotic orbifolds 
matter localization is restricted by the compactification rules. 
All fields of the 4D effective field theories emerging from heterotic compactifications 
arise from the (anomaly, tachyon and ghost free) spectrum of excitations 
of {\it closed} strings that are not affected by the action of the orbifold.

In (supersymmetric) heterotic orbifolds, bulk or {\it untwisted fields} 
correspond to the orbifold--invariant states arising directly from the 10D closed strings of 
the uncompactified heterotic string, whose field limit is 10D $\maN=1$ supergravity endowed 
with an \E8\x\E8 or \SO{32} Yang--Mills theory. Thus, the 4D gauge (super)fields, generating
the unbroken 4D gauge group $\maG_{4D}\subset\E8\x\E8$ or \SO{32}, and some 4D matter
states live in the bulk of a heterotic orbifold.

Additionally, there are the so--called {\it twisted fields}, which arise from strings that 
are closed only due to the action of the orbifold.
Twisted fields are always localized at singularities of the orbifold and are thus
instrumental in the conception of a flavor theory with non--trivial representations from strings.
As long as there are no further compactification ingredients, such as Wilson lines~\cite{Ibanez:1986tp}
or discrete torsion~\cite{Vafa:1986wx,Sharpe:2000ki,Gaberdiel:2004vx,Ploger:2007iq}, that may lead to differences 
in the states at the singular points, the {\it twisted} spectrum is degenerate, i.e. all
singularities carry identical twisted string states.

Couplings among string states are subject to a set of constraints called string 
selection rules~\cite{Hamidi:1986vh,Dixon:1986qv,Casas:1991ac,Kobayashi:1991rp,Kobayashi:2011cw,Nilles:2013lda,Bizet:2013gf},
due to symmetries of the underlying conformal field theory of the compactified string theory.
These selection rules establish for which combination of string states there is a non--zero 
correlation function, and thus a non--zero coupling for the associated effective fields.
In the 4D model emerging from an Abelian heterotic orbifold, the selection rules amount to including additional 
(Abelian $\Z{N}\x\Z{M}\x\cdots$) symmetries and assign thus appropriate discrete 
charges to each field in the model.

Thus, we notice that flavor symmetries in Abelian toroidal heterotic orbifolds 
have two sources: the group of non--Abelian (relabeling) symmetries $G_{nA}$ from the geometrical 
structure of the compactification space and the group of Abelian symmetries $G_A$ 
from the string selection rules.
In the case that the string selection rules provide a normal subgroup (invariant under
conjugation) of the full symmetry group, the resulting flavor symmetry is isomorphic to 
the semi--direct product $G_{nA}\ltimes G_A$ (see e.g.~\cite{Ishimori:2012zz}).

\begin{figure}[!t!]
\begin{center}
\subfigure[Fixed points]{
\label{fig:T2overZ3a}
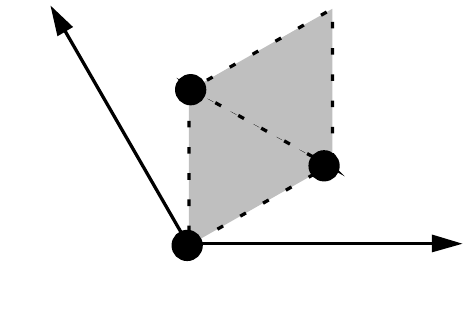
}
\hskip 1cm
\subfigure[Symmetries and charges]{
\label{fig:T2overZ3b}
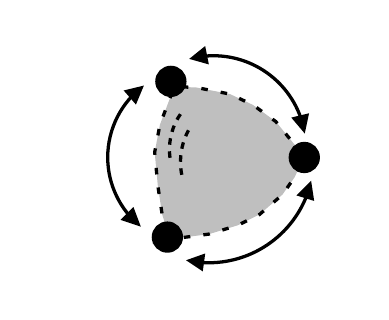
}
\caption{Geometrical origin of a $\Delta(54)$ flavor symmetry in a $\maT^2/\Z3$ orbifold.
If the fixed points are not further affected by the compactification, there is an $S_3$ permutation
symmetry. Further, string selection rules impose additional a $\Z3\x\Z3$ symmetry based on the
localization charges $m$ and $q$ of twisted states. The resulting symmetry is $S_3\ltimes\Z3^2=\Delta(54)$.}
\label{fig:T2overZ3}
\end{center}
\end{figure}

Let us turn now to a relevant example for the present work. Suppose that two extended
dimensions are compactified in the orbifold $\maT^2/\Z3$, where we choose the torus
to be defined by the root lattice $\Lambda_{\SU3}$ which is invariant under the \Z3
generator $\vartheta=e^{2\pi\I/3}$ in complex coordinates. That is, in the orbifold,
points $z_1$ and $z_2$ of $\mathbb C$ are equivalent if they can be related by 
$z_1=\vartheta z_2 + \lambda$, $\lambda\in\Lambda_{\SU3}$. In this orbifold, there exist
three inequivalent fixed points or orbifold singularities\footnote{Analogous results are obtained for 
the second non--trivial \Z3 group element, $\vartheta^2$.}
$z_{f,m},\,m=0,1,2$, such that $z_{f,m}=\vartheta z_{f,m} + \lambda_m$
for some lattice vectors $\lambda_m$. We can choose the inequivalent fixed points to
be $z_{f,0}=0$, $z_{f,1}=\frac13(2 e_1  + e_2)$ and $z_{f,2}=\frac13(e_1  + 2 e_2)$,
where $\{e_\alpha\}$ span $\Lambda_{\SU3}$, as depicted in fig.~\ref{fig:T2overZ3a}. The 
gray region contains all inequivalent points in this orbifold.

Remarking in fig.~\ref{fig:T2overZ3a} that the upper tip is equivalent to $z_{f,0}$ 
and that the lines on both sides of $z_{f,i}$, $i\neq0$, are identified, the orbifold 
becomes the triangular pillow--like object with 
three apices displayed in fig.~\ref{fig:T2overZ3b}. This solid is clearly invariant under
all possible apex permutations, as symbolized by the arrows in that figure. Thus, we identify a
geometrical $S_3$ symmetry.

When the current example is applied to heterotic orbifolds, the string selection rules demand
additionally that any coupling of the form $\Phi_{m_1}\Phi_{m_2}\Phi_{m_3}\cdots$ among string states $\Phi_{m_i}$
(setting $m=0$ for untwisted states) satisfy first $\sum_i m_i = 0\mod 3$. Noting that this relation
corresponds to a \Z3 symmetry, it can be rewritten as $\prod_i \kappa^{m_i}=\mathbbm1$ in terms of
a \Z3 generator $\kappa=e^{2\pi\I/3}$. Furthermore, assigning a charge $q=1$ to $\vartheta$--twisted states 
(and $q=2$ to $\vartheta^2$--twisted states and $q=0$ to untwisted states),
non--vanishing string couplings require that the couplings themselves be non--twisted, 
i.e. $\prod_i\vartheta^{q_i}=\mathbbm1$, which can be rewritten as $\sum_i q_i = 0\mod3$. 
Thus, we identify a \Z3\x\Z3 arising from the selection rules.
Finally, since the \Z3\x\Z3 obtained is a normal subgroup of the group generated by $S_3$ and $\Z3\x\Z3$, then the resulting
effective flavor symmetry of a $\maT^2/\Z3$ orbifold can be written as $\Delta(54)=S_3\ltimes\Z3^2$.

In the absence of Wilson lines and discrete torsion, twisted string states replicate in all 
orbifold singularities, thus appearing always with a multiplicity of three
and building triplet representations. Since $\vartheta^{-1} = \vartheta^2$, twisted
states located at the $\vartheta^2$ fixed points have the opposite geometrical
quantum numbers of the $\vartheta$--twisted states. That is, if we label as $\rep3_{11}$
the $\vartheta$--twisted states,\footnote{We follow here the notation of~\cite{Ishimori:2012zz} 
for $\Delta(54)$ representations; see appendix A.} those generated at the $\vartheta^2$ singularities build then the
representation $\rep3_{12}$. Untwisted states and twisted states affected by Wilson lines or
discrete torsion are just $\Delta(54)$ trivial singlets $\rep1_0$. No other $\Delta(54)$
representations appear in this context, yielding a tight and useful string constraint for 
flavor phenomenology.

This discussion has been explicitly developed for all possible sub--orbifolds (in less than six dimensions) 
appearing in Abelian toroidal heterotic orbifolds~\cite{Kobayashi:2006wq}, resulting in a reduced number 
of family symmetries. The findings include, besides $\Delta(54)$, only the symmetries\footnote{Note
though that, under certain conditions, other symmetries may appear, as in~\cite{Altarelli:2006kg}.}
$D_4$, $(D_4\x D_4)/\Z2$, $(D_4\x\Z4)/\Z2$, $(D_4\x\Z8)/\Z2$ and $S_7\ltimes\Z7^6$.
As we shall see, these symmetries are enlarged in the full 6D heterotic orbifold,
but can then be finally reduced back to these symmetries in phenomenologically viable models.
This may already be considered a phenomenologically relevant observation: not
any flavor symmetry is allowed in particle physics if it arises
from a compactified string theory.

\section{Classification of \Z3\x\Z3 heterotic orbifolds with $\Delta(54)$}
\label{sec:class}

The purpose of this section is to identify string models exhibiting a number of semi--realistic
properties and $\Delta(54)$ flavor symmetry in the simplest compactification scheme where such 
models are present, \Z3\x\Z3 heterotic orbifolds.

\Z3\x\Z3 heterotic orbifolds are characterized by the quotient of
a so--called factorizable torus $\maT^6=\maT^2_1\x\maT^2_2\x\maT^2_3$ divided 
by the joint action of two \Z3 isometries of $\maT^6$
in the extra dimensions of a heterotic string. In the simplest case,\footnote{There
are 15 \Z3\x\Z3 choices, among which many include rototranslations~\cite{Fischer:2012qj}.}
the tori are described by the root lattice of $\SU3_1\x\SU3_2\x\SU3_3$
and the \Z3 generators act diagonally on the tori as
\begin{equation}
  \vartheta = \diag\left(e^{2\pi\I v_1},e^{2\pi\I v_2},e^{2\pi\I v_3}\right)\,,\qquad
  \omega    = \diag\left(e^{2\pi\I w_1},e^{2\pi\I w_2},e^{2\pi\I w_3}\right)\,,
  \label{eq:Z3xZ3twists}
\end{equation}
where $v$ and $w$ are the so--called twist vectors
\begin{equation}
  v = \left(1/3,0,-1/3\right),\qquad w=\left(1/3,-1/3,0\right)\,.
  \label{eq:Z3xZ3twistvectors}
\end{equation}
Consequently, each of the 2--tori are subject to a \Z3 orbifold. 

According to our previous discussion, one may conjecture that these constructions
lead to a $\Delta(54)^3$ flavor symmetry, but this is wrong. In fact, in this
case the relabeling symmetry that naturally appears is $S_3\x S_3\x S_3$.
Further, concerning the symmetries due to string selection rules,
invariance under the two twists, $\vartheta$ and $\omega$, leads to
two \Z3 symmetries analogous to the one for the $q$ charge in the previous section.
In addition, localization selection rules introduce one extra \Z3 factor for each
2--torus. That is, the natural flavor symmetry in these heterotic orbifolds
is $(S_3\x S_3\x S_3)\ltimes\Z3^5$.

Note, however, that the relabeling symmetry can be further enhanced to 
$S_{27}$ if the sizes of the tori $\maT^2_a$, $a=1,2,3$, 
are identical and no Wilson lines nor discrete torsion is invoked. As we shall shortly see, 
phenomenologically viable models only arise if one introduces Wilson lines. In fact,
most promising models have two Wilson lines that distinguish the states located at the singularities 
of two of the tori, retaining only the non-Abelian $S_3$ relabeling symmetry.
Further, there is no reason why all tori should have the same size; their sizes
(and also their shapes) are encoded in the values of (untwisted) moduli that can 
{\it a priori} have arbitrary values.

Once the generic geometrical aspects of the compactification have been set, our
task is now to apply this compactification to a heterotic string. We restrict ourselves
here to the $\maN=1$ \E8\x\E8 heterotic string, but expect similar results from
the $\maN=1$ \SO{32} heterotic string.\footnote{We also expect promising non--supersymmetric
models arising from the $\maN=0$ \SO{16}\x\SO{16} heterotic string, although the
presence of tachyons at some level of the theory would still be a worry.}
Modular invariance of the partition function demands the orbifold to be embedded 
into the gauge group \E8\x\E8. This gauge embedding consists in choosing a 16D 
(shift) vector for each of the twists performed in the six compact dimensions
and a so--called 16D Wilson--line vector $A_\alpha$, $\alpha=1,\ldots,6$, 
encoding in the gauge degrees of freedom each $e_\alpha$ of $\maT^6$. 

The gauge embedding is subject to three constraints. First, 
modular invariance additionally imposes in \Z3\x\Z3 heterotic orbifolds that~\cite{Ploger:2007iq}
\begin{eqnarray}
\label{eq:ModInv}
3\,(V^2-v^2) = 0\mod2\,,           &\quad& 3\,(V\cdot A_\alpha) = 0\mod 2\,,\quad\alpha=1,\ldots,6\,,\\
3\,(W^2-w^2) = 0\mod2\,,           &\quad& 3\,(W\cdot A_\alpha) = 0\mod 2\,, \nonumber\\
3\,(V\cdot W-v\cdot w) = 0\mod2\,, &\quad& 3\  A_\alpha^2 = 0\mod 2\,,\nonumber\\
                                   &\quad& 3\,(A_\alpha\cdot A_\beta) = 0\mod 2\,,\quad \alpha\neq\beta\,,\nonumber
\end{eqnarray}
where $V$ and $W$ are the 16D vectors that denote respectively the gauge embeddings of the twists $v$ and $w$ 
of eq.~\eqref{eq:Z3xZ3twistvectors}. Secondly, both $V$ and $W$ must be consistent
with a \Z3\x\Z3 action. This amounts to requiring that three times these
vectors must be a trivial gauge transformation within \E8\x\E8, i.e.
for the shift vector $V$ (with entries $V^{(i)}$)~\footnote{These constraints
arise from the fact that the root lattice of each \E8 is even (and self--dual). 
An arbitrary shift within the lattice does not alter the gauge degrees of freedom.}
\begin{equation}
3\sum_{i=1}^8 V^{(i)} = 0\mod 2\,,\qquad 3\sum_{i=9}^{16} V^{(i)} = 0\mod 2\,,
\label{eq:lattice-condition}
\end{equation}
demanding that the entries $V^{(i)}$ be all integer or half--integer, independently for $i=1,\ldots,8$
and $i=9,\ldots,16$. Analogous conditions must then be imposed to $W$.
The final constraint imposes that Wilson--line vectors must be consistent with
the choice of $\maT^6$ lattice and the action of the orbifold on it.
The fact that the lattice vectors $e_\alpha$ are related by the action of $\vartheta$ and $\omega$ 
translates to relations among all $A_\alpha$. For instance, in \Z3\x\Z3 heterotic orbifolds 
since $e_2 = \vartheta e_1$ (see e.g. fig.~\ref{fig:T2overZ3a}, valid in this case), 
then $A_1 = A_2$ up to a trivial gauge transformation in \E8\x\E8. 
One finds that these geometrical considerations lead to the conditions
\begin{eqnarray}
\label{eq:condWL}
A_\alpha = A_{\alpha+1}\,, &&\alpha=1,3,5\,,\\
3\sum_{i=1}^8 A_\alpha^{(i)} = 0\mod 2\,,&& 3\sum_{i=9}^{16} A_\alpha^{(i)} = 0\mod 2\,.\nonumber
\end{eqnarray}

A comment is in order. Notice that each 2--torus can be affected by up to one inequivalent, non--trivial
Wilson line. If one includes the Wilson line $A_{2a-1} = A_{2a}$ associated with the compactification
in the $\maT^2_a$ torus, $a=1,2,3$, the relabeling symmetry $S_3$ of that torus disappears.
Thus, with one and two non--vanishing Wilson lines, the non--Abelian relabeling symmetry gets
broken down, respectively, to $S_3\x S_3$ and $S_3$, while no non--Abelian symmetry is left when
all three Wilson lines are non--trivial. Hence, it follows that only models with two non--trivial Wilson lines
can lead to a $\Delta(54)=S_3\ltimes\Z3^2$ flavor symmetry in \Z3\x\Z3 heterotic orbifolds.

After finding solutions to the constraints~\eqref{eq:ModInv}--\eqref{eq:condWL}, there are standard 
techniques, discussed elsewhere in great detail (see e.g.~\cite{RamosSanchez:2008tn,Vaudrevange:2008sm}),
to determine the spectrum of massless string states, including their gauge quantum numbers, localization, 
couplings and other properties of the {\it supersymmetric} effective field theory. Spectra obtained this way must then be
inspected from a phenomenological perspective, imposing criteria based on observable particle physics
(and/or cosmology) that may discriminate phenomenologically viable models from others.

Clearly, given the number of gauge--embedding parameters, the constraints~\eqref{eq:ModInv}--\eqref{eq:condWL} 
can be satisfied for a large number of shift and Wilson--line vectors, making the task of identifying
phenomenologically viable heterotic orbifolds very time--consuming. Fortunately, this task becomes accessible
thanks to tools such as the \texttt{orbifolder}~\cite{Nilles:2011aj}, which automatizes the computation
of massless spectra, couplings and other important features of the models.

With the purpose of finding promising models endowed with a $\Delta(54)$ flavor symmetry, 
we have used the \texttt{orbifolder} to randomly construct a large number of inequivalent \Z3\x\Z3 
heterotic orbifold models. Models are considered to be equivalent by the software if no differences are 
found when comparing the full gauge group, the non-Abelian gauge quantum numbers of the
resulting states and the number of non-Abelian gauge singlets in the massless spectrum.
From the created models, we have then selected the most promising ones. Here,
a promising model must yield the SM gauge group, such that the hypercharge generator be
non--anomalous and (with normalization) compatible with grand unification, three generations 
of quarks and leptons, at least a couple of Higgs (super)fields, $H_u$ and $H_d$, 
and only vectorlike exotics w.r.t. the SM gauge group.

Our results are as follows. We have obtained over $7\x10^{6}$  
inequivalent \Z3\x\Z3 heterotic orbifold models~\footnote{Following the statistical approach 
of~\cite[sec. 2.2]{Lebedev:2008un}, we estimate that the number of generated models represents
about 90\% of the total of possible models in this scenario.}, with up to (the maximum of) three 
inequivalent Wilson lines. After applying our phenomenological constraints, only 
789 models exhibit the required properties. We have verified that, considering 
couplings\footnote{Given the persistent controversy about the selection rules in heterotic orbifolds, 
we have considered only the so--called rule 4~\cite{Kobayashi:2011cw}, gauge and space--group invariance,
and $R$--charge conservation~\cite{Nilles:2013lda,Bizet:2013gf}.}
of the vectorlike exotics with up to six SM singlets, in a large number
of these models all exotics decouple once the SM singlets develop vacuum expectation values (VEVs). 
Other models require higher dimensional operators to yield mass terms for all vectorlike exotics.

An interesting geometrical quality of the promising models regards the 
effective family symmetry. Among the 789 selected models, most (696) of them  
have two inequivalent non--vanishing Wilson lines. About 10\% of the 
viable models (81 of them) require one non--trivial Wilson line, 
and only 12 result from compactifications with three Wilson lines. 
Therefore, we find that $\Delta(54)$ as a flavor symmetry of (MS)SM--like models 
is favored in \Z3\x\Z3 heterotic orbifold models.

This outcome is compatible with previous results found in the 
literature. Particularly, in ref.~\cite{Nilles:2014owa} the authors have found 
445 \Z3\x\Z3 heterotic orbifold models with the properties we have required, 
out of which 369 of them exhibit two non--trivial Wilson lines. In this
perspective, our search shows to be more exhaustive.

\section{A sample model with stringy $\Delta(54)$ flavor}
\label{sec:stringyDelta54}

With the purpose of exploring the flavor phenomenology produced by string compactifications,
let us now study the properties of one of the promising models 
from our \Z3\x\Z3 heterotic orbifold scan, chosen due to its simplicity.
The parameters that define the model are the shift vectors
\begin{subequations}
\label{eqs:shifts_model9}
\begin{eqnarray}
  3 V &=& \left(-\tfrac{1}{2}, -\tfrac{1}{2}, -\tfrac{1}{2}, -\tfrac{1}{2}, \tfrac{1}{2}, \tfrac{1}{2}, \tfrac{1}{2}, \tfrac{1}{2}; -2, 0, 0, 1, 1, 1, 1, 4\right)\,, \\
  3 W &=& \left( 0, 1, 1, 4, 0, 0, 1, 1; 1, -1, 4, -4, -1, 0, 0, 1\right)\,,
\end{eqnarray}
\end{subequations}
and the Wilson lines
\begin{subequations}
\label{eqs:WL_model9}
\begin{eqnarray}
  3 A_{1} = 3 A_2 & = & \left(-\tfrac{7}{2}, -\tfrac{3}{2}, \tfrac{9}{2}, \tfrac{7}{2}, -\tfrac{7}{2}, -\tfrac{3}{2}, \tfrac{5}{2}, \tfrac{7}{2};  -3, 0, -2, 0, -2, -4, 3, -2\right)\,,\\
  3 A_{3} = 3 A_4 & = & \left( 3, 3, -3, -2, -1, 2, 4, -4;  -3, 1, -1, -4, 1, 1, 4, 1\right)\,.
\end{eqnarray}
\end{subequations}
These parameters yield the unbroken gauge group $\SU3_C\x\SU2_L\x\U1_Y\x[\SU2\x\U1^{11}]$, where the
additional \SU2 factor is considered hidden because no SM--field carries a charge under that group.
However, all fields in the spectrum are charged under the additional \U1 factors.

\begin{table}[!t!]
\begin{center}
\begin{tabular}{|rlcc|rlcc|}
\hline
 \#  &  irrep & $\Delta(54)$ & label & \# & anti-irrep  & $\Delta(54)$ & label \\
\hline
3 & $\left(\rep{3}, \rep{2}\right)_{\frac{1}{6}}$ & $\rep3_{11}$ & $Q_i$ &  & & & \\
3 & $\left(\crep{3}, \rep{1}\right)_{-\frac{2}{3}}$ & $\rep3_{11}$ & $\bar{u}_i$ &  & & & \\
3 & $\left(\crep{3}, \rep{1}\right)_{\frac{1}{3}}$ & $\rep3_{11}$ & $\bar{d}_i$ &  & & & \\
3 & $\left(\rep{1}, \rep{2}\right)_{-\frac{1}{2}}$ & $\rep3_{11}$ & $L_i$ &  &  & &  \\ 
3 & $\left(\rep{1}, \rep{1}\right)_{1}$ & $\rep3_{11}$ & $\bar{e}_i$   &  & & & \\
3 & $\left(\rep{1}, \rep{1}\right)_{0}$ & $\rep3_{12}$ & $\bar{\nu}_i$ &  & & & \\
1 & $\left(\rep{1}, \rep{2}\right)_{-\frac{1}{2}}$ & $\rep1_0$ & $H_d$ & 1 & $\left(\rep{1}, \rep{2}\right)_{\frac{1}{2}}$ & $\rep1_0$ & $H_u$ \\ 
\hline
\multicolumn{8}{c}{Flavons}\\
\hline
3 & $\left(\rep{1}, \rep{1}\right)_{0}$ & $\rep3_{11}$ & $\phi^{u}_i$   &  & & & \\
3 & $\left(\rep{1}, \rep{1}\right)_{0}$ & $\rep3_{11}$ & $\phi^{d,e}_i$ &  & & & \\
3 & $\left(\rep{1}, \rep{1}\right)_{0}$ & $\rep3_{12}$ & $\bar\phi^{\nu}_i$ & & & & \\
2 & $\left(\rep{1}, \rep{1}\right)_{0}$ & $2\p\rep1_0$ & $s^{(d,e)},s^u$ &  & & & \\
128 & $\left(\rep{1}, \rep{1}\right)_{0}$ & $77\p\rep1_0+16\p\rep3_{12}+\rep3_{11}$ & $N_i$ &  &  & &  \\ 
\hline
\multicolumn{8}{c}{Exotic states}\\
\hline
16 & $\left(\rep{1}, \rep{2}\right)_{\frac{1}{6}}$ & $10\p\rep1_0+2\p\rep3_{12}$ & $v_i$  & 16 & $\left(\rep{1}, \rep{2}\right)_{-\frac{1}{6}}$ & $4\p\rep1_0+4\p\rep3_{12}$ & $\bar v_i$ \\
3  & $\left(\rep{3}, \rep{1}\right)_{0}$ & $\rep3_{12}$ & $y_i$                           & 3 & $\left(\crep{3}, \rep{1}\right)_{0}$ & $3\x\rep1_0$ & $\bar y_i$ \\
1  & $\left(\crep{3}, \rep{1}\right)_{-\frac{1}{3}}$ & $\rep1_0$ & $z_i$                  & 1 & $\left(\rep{3}, \rep{1}\right)_{\frac{1}{3}}$ & $\rep1_0$ & $\bar z_i$ \\
7  & $\left(\rep{1}, \rep{1}\right)_{-\frac{2}{3}}$ & $4\p\rep1_0+\rep3_{12}$ & $x_i$     & 7 & $\left(\rep{1}, \rep{1}\right)_{\frac{2}{3}}$ & $4\p\rep1_0+\rep3_{11}$ &  $\bar x_i$ \\
51 & $\left(\rep{1}, \rep{1}\right)_{-\frac{1}{3}}$ & $30\p\rep1_0+7\p\rep3_{12}$ & $w_i$ & 51 & $\left(\rep{1}, \rep{1}\right)_{\frac{1}{3}}$ & $24\p\rep1_0+9\p\rep3_{12}$ & $\bar w_i$ \\
\hline
\end{tabular}
\caption{Massless spectrum. Representations w.r.t. $\SU3_C\times\SU2_L$ 
are given  in  bold face, the hypercharge is indicated by the subscript. 
The 3rd and 7th columns display the $\Delta(54)$ flavor representations.}
\label{tab:spectrum_model9}
\end{center}
\end{table}

\begin{table}[!h!]
\begin{center}
\begin{tabular}{|c|c|c|c|c|c|c||c|c||c|c|c||c|c|}
\hline
$\phantom{A^{A^{A^A}}}$& $Q_i$ & $\bar{d}_i^c$ & $\bar{u}_i^c$ & $L_i$ & $\bar{e}_i^c$ &$\bar{\nu}_i$& $H_u$ & $H_d$ & $\phi^u_i$ & $\phi^{(d,e)}_i$ & $\bar{\phi}^\nu_i$ & $s^u$ & $s^{(d,e)}$\\
\hline
$\Delta(54)$ &$\bs3_{11}$ & $\bs3_{11}$  & $\bs3_{11}$  & $\bs3_{11}$  & $\bs3_{11}$&  $\bs3_{12}$  & $\bs1_0$ & $\bs1_0$  & $\bs3_{11}$ & $\bs3_{11}$ & $\bs3_{12}$ & $\bs1_0$   & $\bs1_0$ \\
\hline
$\Z3^{(1)}$  &  $\omega$  & $1$          & $\omega$     & $1$          & $\omega$   &  $1$          & $1$      & $1$       & $1$         & $1$         & $1$         & $\omega$   & $\omega^2$ \\
\hline
$\Z3^{(2)}$  & $1$        & $\omega^2$   & $1$          & $\omega^2$   & $1$        &  $\omega$     & $1$      & $1$       & $1$         & $\omega$    & $\omega$    & $1$        & $1$ \\
\hline
$\Z3^{(3)}$  & $\omega$   & $1$          & $\omega$     & $1$          & $\omega$   &  $1$          & $1$      & $1$       & $1$         & $1$         & $1$         & $\omega$   & $\omega^2$ \\
\hline
\end{tabular}
\caption{Flavor representations for the SM matter and flavon fields in a \Z3\x\Z3 sample model. 
The \Z3 charges are defined in terms of the twist and field localizations in~\eqref{eq:Z3generators} with $\omega=e^{2\pi\I/3}$.}
\label{tab:matterfields}
\end{center}
\end{table}

Due to its two Wilson lines~\eqref{eqs:WL_model9}, the model has the flavor symmetry $S_3\ltimes\Z3^5\supset\Delta(54)$. 
The $\Delta(54)$ quantum numbers are associated with the symmetries of the third torus, $\maT^2_3$, whose
localized states are not affected by any Wilson line. If we allow for the spontaneous breakdown of the 
three additional \Z3 symmetries by VEVs of appropriate SM singlets transforming as $\bs1_0$ under $\Delta(54)$, 
the flavor symmetry in the vacuum is just $\Delta(54)$ and the extra $[\SU2\x\U1^{11}]$ gauge factors are broken too.

The gauge and $\Delta(54)$ representations of the massless matter spectrum of our sample model
are provided in table~\ref{tab:spectrum_model9}. As explained before, the only possible
$\Delta(54)$ representations are the trivial singlet $\rep1_0$ and the two triplets $\rep3_{11}$ 
and $\rep3_{12}$. One particular feature of the observable sector is that, directly from the string computation, 
only three SM generations that build non--trivial flavor representations arise, while the 
Higgs states are untwisted fields and thus uncharged under the flavor symmetry. On the other hand,
the exotic particles are vectorlike w.r.t. the SM gauge group, but not necessarily under the flavor group. 
Despite this hurdle, there exist SM singlets $N_i$ in the appropriate flavor representations, so 
that all exotics and the singlets $N_i$ themselves can acquire masses when $\vev{N_i}\neq0$.

To understand better the flavor phenomenology of the observable sector of this model,
we display in table~\ref{tab:matterfields} all flavor charges of the SM superfields and some gauge 
singlets that shall serve as flavons. The \Z3 charges are given in terms of
\begin{equation}
\label{eq:Z3generators}
 \omega^q\qquad\text{ and }\qquad \kappa_{a}^{\ m_a}\equiv (e^{2\pi\I/3})^{m_a}\,\qquad
 \text{with}\quad a=1,2;\,q,m_a=0,1,2\,,
\end{equation}
where $\omega$ is the (eigenvalue of the) second twist in eq.~\eqref{eq:Z3xZ3twists}, $\kappa_a$ correspond to the \Z3 
generators associated with the localization labels $m_a$ in the (first or second) torus  $\maT^2_a$, 
as described in section~\ref{sec:symmetries}, and $q$ is the power of the twist that yields 
the corresponding twisted states. Note that $\omega=\kappa_a=e^{2\pi\I/3}$.

Since the SM matter fields are charged under flavor symmetries, the presence of the properly charged
$s$ and $\phi$ flavon fields allows for Yukawa couplings in the (non--renormalizable) superpotential, 
which in this case can be written as follows
\begin{eqnarray}
\label{eq:Wyuk}
W_{Y} &=& y_{ijk}^u Q_i H_u \bar{u}_j \phi^u_k s_u + y_{ijk}^d Q_i H_d \bar{d}_j \phi^{(d,e)}_k s^{(d,e)} + y_{ijk}^e L_i H_d \bar{e}_j \phi^{(d,e)}_k s^{(d,e)} \\
      &+& y_{ijkl}^\nu L_i H_u \bar{\nu}_j+ \lambda_{ijk} \bar\nu_i \bar\nu_j \bar\phi^\nu_k\,, \qquad\qquad\qquad\qquad i,j,k=1,2,3,  \nonumber
\end{eqnarray}
where the summation over repeated indices must follow the rules of the product of $\Delta(54)$ representations
that lead to invariant singlets (cf. appendix A),
$$\rep1_0\subset \rep3_{11}\x\rep3_{12},\quad \rep1_0\subset \rep3_{11}\x\rep3_{11}\x\rep3_{11}\,,\quad \rep1_0\subset \rep3_{12}\x\rep3_{12}\x\rep3_{12}.$$

In principle, all Yukawa--coupling coefficients, $y$ and $\lambda$, are computable by applying CFT techniques for the string model.
However, it is known that there are still some challenges to be solved for non--renormalizable couplings.
The best we can do here is to estimate that $y$ are order one (but with a suppression due non--renormalizability) 
because they include the untwisted Higgs fields, whereas $\lambda$ must be somewhat suppressed because all involved 
fields are twisted. We observe that the second row of $W_Y$ admits neutrino masses from a type I see--saw
mechanism with three right--handed (RH) neutrinos with proper $\phi$ flavon VEVs. 
Similarly, the Dirac masses of charged leptons and quarks are determined by the VEVs of other
flavons $\phi$ and $s$. We point out that the structure of masses for down--quarks and charged leptons
is predicted in this model to be identical because the flavons involved in the corresponding couplings
are unavoidably the same. As we shall see, this enforces a more stringent sort of $b-\tau$ unification.

\section{Fermion masses from a $\Delta(54)$ flavor symmetry}
\label{sec:Delta54pheno}

The properties of the string--derived model presented before can be now studied
from a bottom--up perspective. Although our string sample model is supersymmetric and 
all couplings are determined at the compactification scale, the general structure 
of the Yukawa Lagrangian at low energies can be determined from $W_Y$, if we insist on retaining
the $\Delta(54)$ flavor symmetry in the soft--breaking sector. Besides, it
is known that Yukawa couplings do not receive large contributions 
through the renormalization running~\cite{Xing:2007fb}. Similarly, threshold 
corrections shall not alter the mass and mixing structure of quarks and leptons,
since it depends mainly on mass ratios.
Therefore, we can safely study the viability of the model by restricting ourselves
to the behavior of the appropriate non--supersymmetric fields.

In a compact notation, the effective Yukawa Lagrangian for quarks and charged leptons 
that is obtained from $W_Y$ reads
\begin{eqnarray}
\label{eq:Lyuk}
{\cal L}_{Y}^f&=&y_1^f \left[F_1  H\bar{f}_1 \phi_1+F_2 H\bar{f}_2   \phi_2+F_3 H\bar{f}_3   \phi_3\right] \\
&+&y_2^f\left[(F_1  H\bar{f}_2+F_2  H\bar{f}_1)\phi_3+(F_3  H\bar{f}_1+F_1  H\bar{f}_3)\phi_2+(F_2 H \bar{f}_3+F_3  H\bar{f}_2)\phi_1\right]+ h.c.\,,\nonumber
\end{eqnarray}
where generically $F$ and $\bar f$ denote respectively the left--chiral and right-chiral components of SM fermions, 
$H$ labels the Higgs associated with $\bar f$, and $\phi$ stands for flavon fields. 
Further, we have let the VEVs of the $s$ flavons be absorbed in the Yukawas $y$, as
they do not alter the structure of the couplings.

From the Yukawa Lagrangian~\eqref{eq:Lyuk}, the Dirac mass matrices 
for the charged fermions (namely, up and down quarks, and charged leptons) 
generically take the form
\begin{equation}\label{diracmass}M_f^D=
\left(\begin{array}{ccc}
y_1^f \phi_1^f&y_2^f \phi_3^f&y_2^f \phi_2^f\\
y_2^f \phi_3^f&y_1^f \phi_2^f&y_2^f \phi_1^f\\
y_2^f \phi_2^f&y_2^f \phi_1^f&y_1^f \phi_3^f
\end{array}
\right)\,.
\end{equation}

Let us now make a phenomenological assumption on the flavon VEVs. Suppose the possibility of 
a VEV alignment of the form $\langle \phi^f \rangle= v_{\phi}^f(0,r^f,1)$, 
with $f=u,d,e$, for some real values $v_\phi^f$ and $r^f$.
This greatly simplifies the mass matrices to
\begin{equation}\label{mass:charged1}M_f^D= 
\left(\begin{array}{ccc}
0&a^f&a^f r^f\\
a^f &b^f r^f&0\\
a^f r^f&0&b^f
\end{array}
\right)\,,
\end{equation}
where we define $a^f\equiv y_2^f v_{\phi}^f$ and $b^f\equiv y_1^f v_{\phi}^f$.
Using now the invariant traces and determinant of $M_f^D$ (we take a negative 
$m_1^f$ to compensate the minus sign in the determinant),
\begin{eqnarray}
\label{eq:Minvariants}
\tr M^D_f    &=&  b^f\,(1+r^f) \hspace{27mm}              \stackrel{!}{=} -m^f_1+m^f_2+m^f_3\,, \\
\tr(M^D_f)^2 &=& [2(a^f)^2+(b^f)^2][1+(r^f)^2]\,          \stackrel{!}{=} \ (m_1^f)^2+(m_2^f)^2+(m_3^f)^2\,, \nonumber\\
\det M^D_f   &=& -(a^f)^2 b^f\,[1+(r^f)^3] \hspace{11mm} \stackrel{!}{=} -m^f_1 m^f_2 m^f_3\,, \nonumber
\end{eqnarray}
it is straightforward to write down the Dirac mass matrices in terms of its eigenvalues, i.e.
the three (observable) fermion masses of type $f$, $m_i^f$.

Clearly, any solution to the invariants~\eqref{eq:Minvariants} provides the right masses
for quarks and charged leptons. If we take e.g. the hierarchical solution, 
i.e. $r^f\ll1$ and $a^f\ll b^f$, the mass matrices take the form
\begin{equation}\label{mass:charged2}
M_f^D\approx
\left(\begin{array}{ccc}
0&\sqrt{m^f_1 m^f_2}& \frac{m^f_2-m^f_1}{m^f_3}\sqrt{m^f_1 m^f_2}\\
\sqrt{m^f_1 m^f_2}&m^f_2-m^f_1&0\\
 \frac{m^f_2-m^f_1}{m^f_3}\sqrt{m^f_1 m^f_2}&0&m^f_3
\end{array}
\right),
\end{equation}
which corresponds to
\begin{equation}
\label{eq:hierarchysol}
r^f \approx (m_2^f-m_1^f) / m_3^f\,,\qquad (a^f)^2\approx m_1^f m_2^f\,,\qquad b^f\approx m_3^f\,.
\end{equation}
We notice that the hierarchical solution is compatible with the hierarchy of observed fermion masses.

In the down--quark sector, this structure gives the Gatto-Sartori-Tonin 
formula for the Cabibbo angle, which is approximately the ratio  
$(M_d^{D})_{12}/(M_d^{D})_{22}$,
\begin{equation}
\label{gst}
\lambda_C\approx \sqrt{\frac{m_d}{m_s}}\,,
\end{equation}
where we additionally used that $m_d/m_s\ll1$.\footnote{Eq.~\eqref{gst} has a small 
correction of order $\sqrt{m_u/m_c}$ from the up--quark sector.}
The other two mixing angles are very small at leading order, 
but could be generated if some of the vectorlike quarks mix with 
the SM quarks, see for instance~\cite{Morisi:2013eca}. 

For charged leptons, on the other hand, the same  flavon VEV alignment must be imposed 
because down--quarks and charged leptons share the same flavons. It follows that 
the corresponding mass matrix is diagonalized by a rotation in the 1--2 entries with 
the mixing angle of the order $\sqrt{m_e/m_\mu}$. 

\begin{figure}[!b!]
\centering
\includegraphics[scale=0.5]{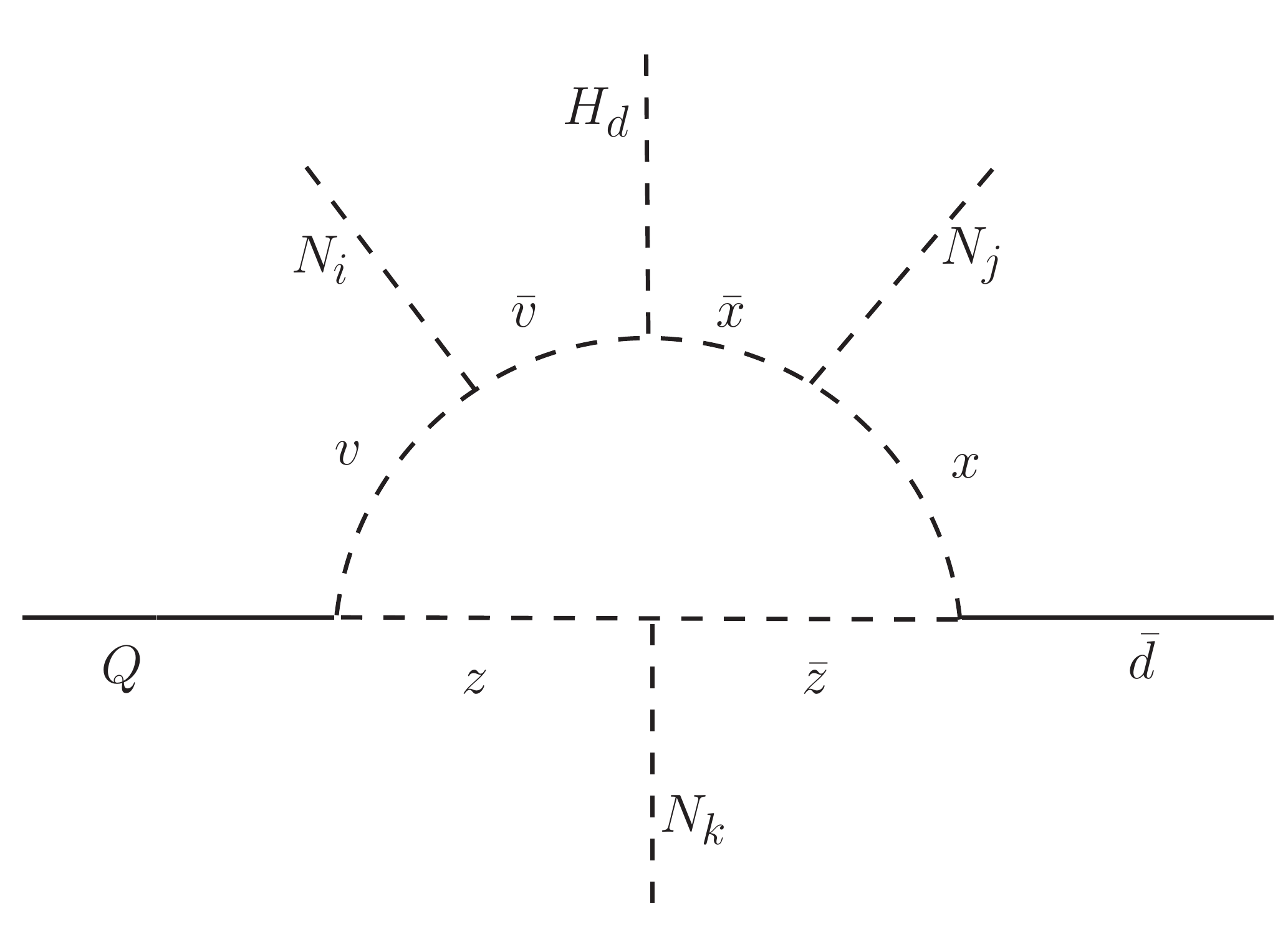}
\includegraphics[scale=0.5]{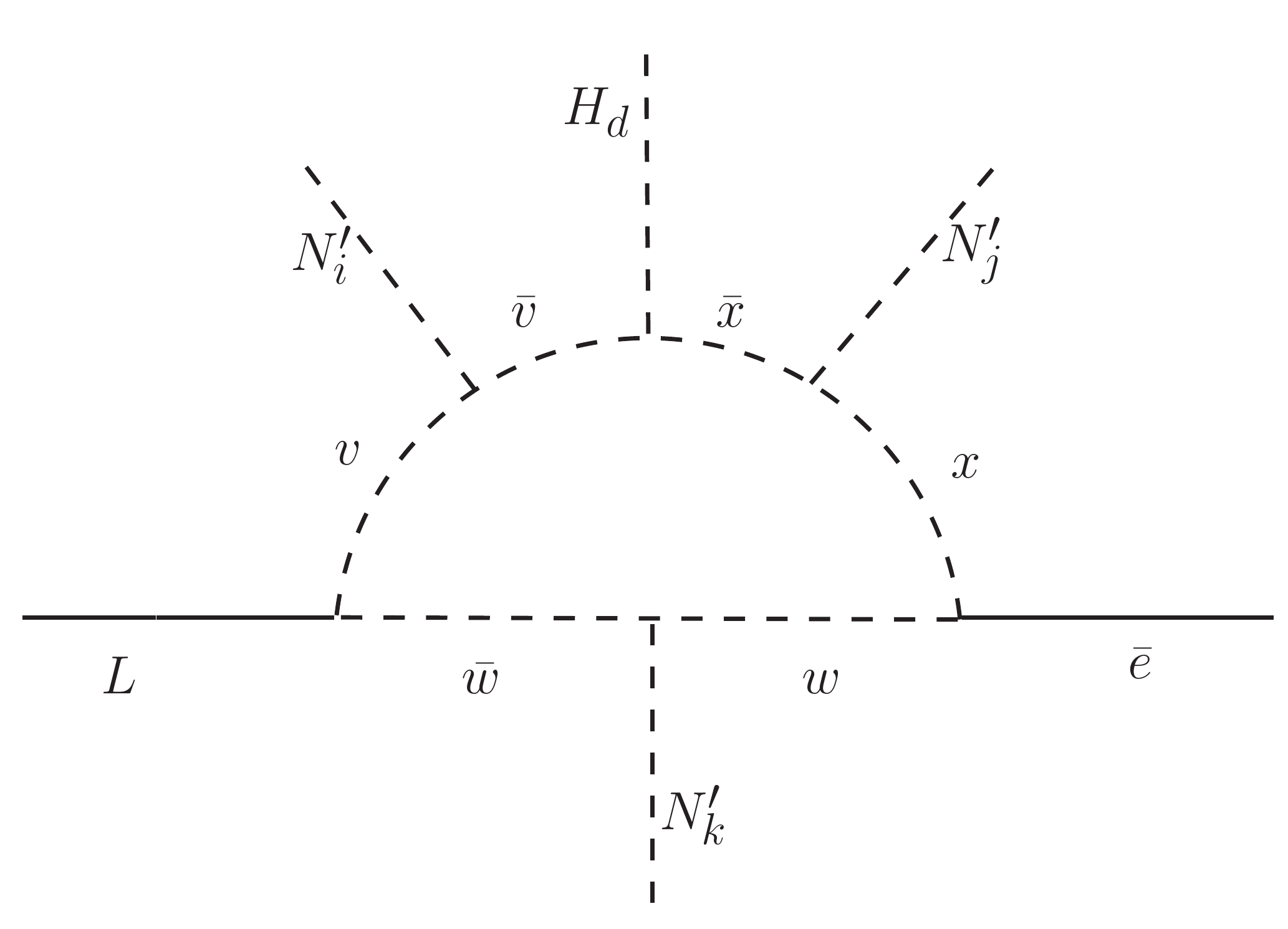}
\caption{Phenomenologically viable operators in the model presented that may alleviate the tension 
observed by the predicted relation~\eqref{eq:massrelation}.}
\label{fig:operatorsvsbottom-tau}
\end{figure}

There is another consequence of the parallelism between the down--quarks and
charged leptons. Since $r^d = r^e$, it follows from eq.~\eqref{eq:hierarchysol}
that the following mass relation in our model is required
\begin{equation}
\label{eq:massrelation}
\frac{m_s-m_d}{m_b} \ \stackrel{!}{=} \ \frac{m_\mu-m_e}{m_\tau}.
\end{equation}
This relation does not match observations. 
We find that some possibilities to amend eq.~\eqref{eq:massrelation} include either to abandon
the flavor structure in the soft-terms of the supersymmetry breaking sector
or that some (colored and uncolored) exotics acquire masses after the breakdown of $\Delta(54)$,
providing different suppression factors for down--quarks and charged leptons.
The latter can be achieved by allowed couplings as those represented in fig.~\ref{fig:operatorsvsbottom-tau},
which yield effective contributions to Yukawa couplings, such as
$$\frac{1}{m_vm_xm_z} Q H_d \bar d \langle N_i N_j N_k\rangle + \frac{1}{m_vm_xm_w} L H_d \bar e \langle N'_i N'_j N'_k \rangle\,,$$
where both $N_{i,j,k}$ and $N'_{i,j,k}$ denote some of the 128 flavons of table~\ref{tab:spectrum_model9}, and $m_{\chi}$ denotes
the effective mass of a given exotic field $\chi$. Realizing particularly that $m_z$ and $m_w$ differ in general and, moreover, that the 
flavons in the couplings may be different, we find that the issue underlined by the constraint~\eqref{eq:massrelation} may be alleviated.
Unfortunately, even if this hurdle is tackled, we do not expect these effects to alter the smallness
of the remaining two quark mixing angles since that depends on the 
hierarchical structure of the fermion masses.

\subsection{The neutrino sector}

For neutrinos, the major difference w.r.t. the other sectors is that, besides 
the presence of Majorana mass terms, neutrinos build a conjugate $\Delta(54)$
triplet, $\rep3_{12}$. Therefore, renormalizable Yukawa couplings become possible.

As stated before, the neutrino masses arise from a type I see--saw according
to the second row of the superpotential~\eqref{eq:Wyuk}.
From there, we can read off the Yukawa Lagrangian for neutrinos:
\begin{eqnarray}
\label{eq:Lyuknu}
{\cal L}_{Y}^\nu&=&y_1^\nu  \left[L_1 H_u\bar{\nu}_1+L_2 H_u\bar{\nu}_2+L_3 H_u\bar{\nu}_3\right] \\
&+&\lambda_1 \left[\bar{\nu}_1 \bar{\nu}_1 \bar\phi_1^\nu + \bar{\nu}_2 \bar{\nu}_2 \bar\phi_2^\nu + \bar{\nu}_3 \bar{\nu}_3\bar\phi_3^\nu\right] \nonumber\\
&+&\lambda_2\left[2\bar{\nu}_1 \bar{\nu}_2 \bar\phi_3^\nu + 2\bar{\nu}_1 \bar{\nu}_3\bar\phi_2^\nu + 2\bar{\nu}_2 \bar{\nu}_3\bar\phi_1^\nu\right]. \nonumber
\end{eqnarray}
Hence, the Dirac neutrino mass matrix is proportional 
to the identity matrix, while RH neutrino masses are governed by a structure 
similar to the one in eq.~\eqref{diracmass}, that is,
\begin{equation}M_{RH}=
\left(\begin{array}{ccc}
\lambda_1 \bar\phi_1^\nu&\lambda_2 \bar\phi_3^\nu&\lambda_2 \bar\phi_2^\nu\\
\lambda_2 \bar\phi_3^\nu&\lambda_1 \bar\phi_2^\nu&\lambda_2 \bar\phi_1^\nu\\
\lambda_2 \bar\phi_2^\nu&\lambda_2 \bar\phi_1^\nu&\lambda_1 \bar\phi_3^\nu
\end{array}
\right).
\end{equation}

We can now make a working assumption about the VEV of the neutrino flavon $\bar\phi^\nu$. 
Considering the alignment $\vev{\bar\phi^\nu}=v_{\nu_3}\left(R_1,\delta,1\right)$,  
the light neutrino mass matrix becomes
\begin{equation}
\label{numass}
  M_\nu~=~\lambda \left(
         \begin{array}{ccc}
              \delta-R^2 R_1^2     & R(-1+R R_1 \delta) & R(-\delta^2+R R_1 )\\
              R(-1+R R_1 \delta)   & R_1-R^2\delta^2    & R(R \delta-R_1^2)\\
              R(-\delta^2+R R_1 )  & R(R \delta-R_1^2)  & R_1\delta-R^2
         \end{array}
         \right),
\end{equation}
where we used the definitions
\begin{equation}
R= {\lambda_2}/{\lambda_1}\,,\qquad
\lambda={y_1^2 \langle H_u\rangle^2}\Big/\left[\lambda_1v_{\nu_3}\left(R_1\delta+2R^3R_1\delta-R^2(1+R_1^3+\delta^3)\right)\right]\,.
\end{equation}
After performing a scan of our parameters, restricting the values of the computed
$\Delta m^2_{12}$, $\Delta m^2_{13}$ and neutrino mixing angles to lie within
the $3\sigma$ region of the global fits~\cite{Forero:2014bxa}, we find that
the mass matrix in eq.~\eqref{numass} is compatible only with a normal hierarchy 
of neutrino masses, i.e. an inverted hierarchy is disfavored, coinciding with 
recent preliminary results from the T2K collaboration~\cite{T2K}.

\begin{figure}[!t!]
\centering
\includegraphics{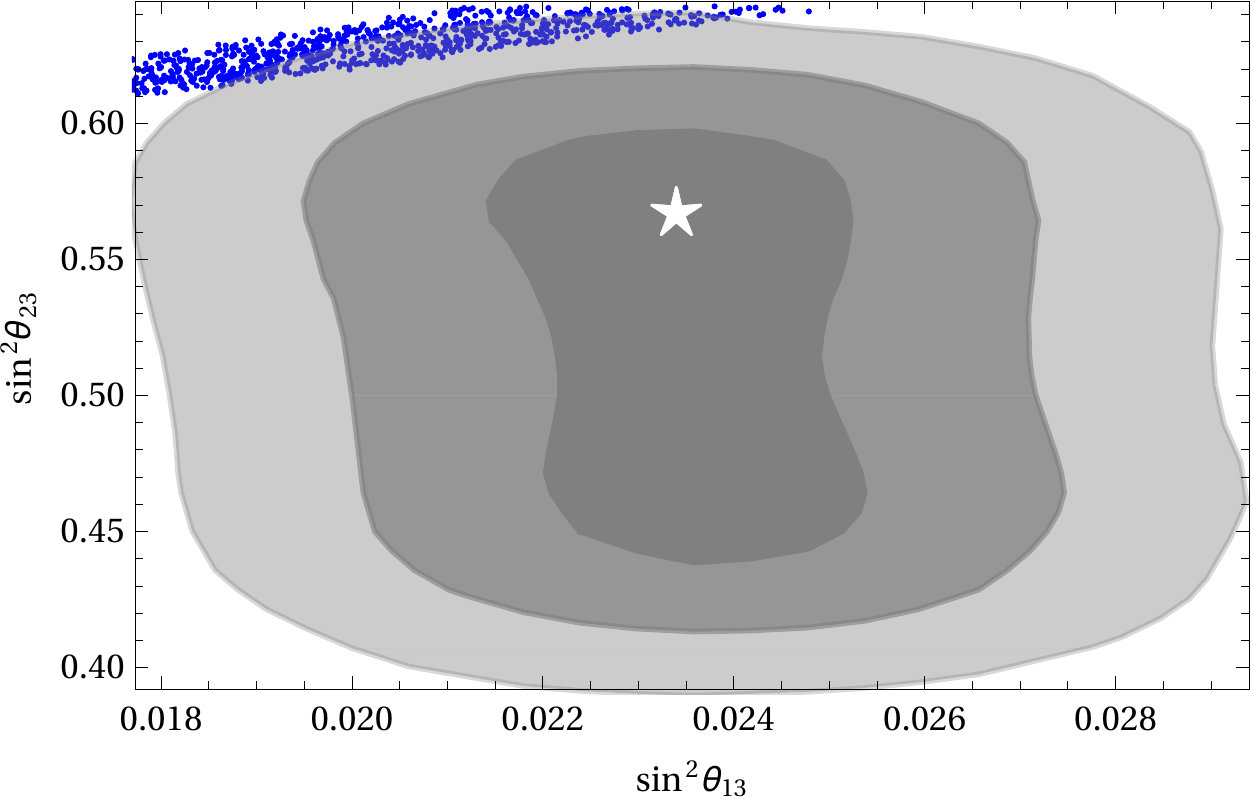}
\caption{Correlation between the atmospheric and reactor mixing angles for normal mass ordering in 
a string--inspired $\Delta(54)$ flavor model. The correlation (blue) points in the upper--left part of 
the plot result from a scan of our parameters $\lambda,\delta,R,R_1$, imposing consistency within
$3\sigma$ with measured values of $\Delta m^2_{12}$, $\Delta m^2_{13}$ and $\theta_{12}$.
The dark/light/lighter gray areas correspond to $1\sigma$/$2\sigma$/$3\sigma$ experimental precision 
around the best fit value (denoted by the star) for the neutrino mixing angles~\cite{Forero:2014bxa}.
}
\label{fig:atmoreactor}
\end{figure}

Furthermore, we observe that our model leads to a correlation between the atmospheric  
and the reactor mixing angles in normal ordering, as displayed by the blue 
region in fig.~\ref{fig:atmoreactor}. Comparing with the precision intervals,
we see that the atmospheric mixing angle lies in the second octant, approximately 
between $51.3$ and $53.1$ degrees, while the reactor mixing angle has values between 
$7.8$ and $8.9$ degrees, in agreement with the oscillation global fits within $3\sigma$. 
These values are crucial for the model since a better measurement of the neutrino 
mixing angles could falsify it.

A final result from our parameter scan is that the lightest neutrino mass, $m_{\nu_1}$, 
takes values in the region between $6$~meV and $6.8$~meV, and the sum of the light 
neutrino masses, $\sum m_{\nu}$, lies in the interval between $65$~meV and $70$~meV,
in consistency with data.

\section{Final remarks}
\label{sec:conclusions}

Flavor symmetries arise naturally in string compactifications, which provide 
a promising ultraviolet completion of usual bottom--up setups. Particularly, 
we have shown that $\Delta(54)$, as a flavor symmetry, appears most naturally 
in semi--realistic \Z3\x\Z3 heterotic orbifold compactifications. We have 
identified almost 700 models with that flavor symmetry and other promising 
particle--physics features, such as SM gauge group and three generations of 
matter fields. By their nature, these constructions reduce the arbitrariness of
low--energy models by constraining the fields and their (flavor and gauge) 
transformation properties and thereby providing useful guidelines
to inspect flavor phenomenology.

To test the viability of $\Delta(54)$ flavor scenarios arising from strings, we 
have studied the phenomenology of one simple string model from our classification,
whose properties may differ from the other identified models. 
In this model, SM fermion fields transform as triplets of the flavor symmetry 
while the Higgs fields do not transform. As a result of the flavor quantum numbers, 
the quarks and charged leptons acquire masses through dimension--6 operators, 
and the Dirac neutrino masses as well as the RH Majorana neutrino masses are generated 
at renormalizable level.  
Furthermore, we observe that choosing some special flavon--VEV alignments results
in the following flavor phenomenology features:
\begin{itemize}
 \item correct masses for quarks and charged leptons;
 \item proper Gatto-Sartori-Tonin relation in the quark sector (although the other two mixing angles are very small);
 \item a mass relation between the down--quark sector and the charged leptonic sector (see eq.~\eqref{eq:massrelation});
 \item compatibility (only) with normal hierarchy of neutrino masses;
 \item smallest neutrino mass of order $6-7$ meV; and
 \item PMNS matrix compatible with current constraints (atmospheric and reactor 
       mixing angles are in the $3\sigma$ region of the global best fit), 
       with the atmospheric mixing angle greater than $45$ degrees.
\end{itemize}
Interestingly, an inverted hierarchy being disfavored as well as the atmospheric 
mixing angle lying in the second octant, are features compatible with recent
preliminary findings of the T2K collaboration~\cite{T2K}.
This outcome lets us assert that \Z3\x\Z3 heterotic orbifolds and $\Delta(54)$ as a flavor 
symmetry provide a fertile playground for useful phenomenology which should be further investigated.

The particular model we have studied here was chosen due to its neat simplicity: 
it has only three SM generations, the extra gauge sector includes only a hidden \SU2 and Abelian symmetries,
and all SM fields build $\Delta(54)$ triplets. These properties are only shared by three more models in the 
set of promising \Z3\x\Z3 compactifications. Other models include additional (exotic) vectorlike pairs of quarks and 
leptons, larger Abelian and non--Abelian hidden gauge symmetries, and some SM fields may build only trivial 
representations of $\Delta(54)$.
This does not imply that other models are more or less promising, but their analysis is somewhat more involved
and shall be the purpose of future studies.

Despite these encouraging features, there are still some challenges to overcome. First,
in heterotic orbifolds it is challenging to obtain the VEV alignments chosen in section~\ref{sec:Delta54pheno}
because VEVs must be settled by a moduli stabilization mechanism that is not fully understood.
Secondly, we found that two of the quark mixing angles in our model are too small and the mass 
relation eq.~\eqref{eq:massrelation} is incorrect. To attempt to alleviate these issues, 
one should study in detail the soft--terms and other corrections in this kind of models. 
Another potential hurdle is the absence of a symmetry that forbids rapid proton decay. 
However, it is conceivable that such symmetry does appear as one of the extra \Z3 symmetries
of another model where matter fields have the correct charges. Finally, as in most flavor 
models, flavor--changing neutral currents pose a challenge that must and shall be studied
elsewhere in the context of our proposal.

\section*{Acknowledgments}
We would like to thank P.K.S. Vaudrevange for useful discussions.
B.~C-P. would like to thank the CLAF-ICyTDF for support during her posdoctoral fellowship.
E.~P. is partly supported by PAPIIT IA101516 and PAPIIT IN111115.
S.~R-S. would like to thank the ICTP for the kind hospitality and support received
through its Junior Associateship Scheme during the realization of this work.

\begin{appendix}
\section{$\Delta(54)$ tensor product for triplet representations}
\label{sec:appendix}

In this appendix, we provide the features of $\Delta(54)$ that are relevant for our proposal, following the
notation of ref.~\cite{Ishimori:2012zz}. The $\Delta(54)$ symmetry group has two one--dimensional, four 
two--dimensional and four three--dimensional irreducible representations. These representations are denoted
as $\rep1_0$ (invariant under the group), $\rep1_{1}$, $\rep2_1$, $\rep2_2$, $\rep2_3$, 
$\rep2_4$, $\rep3_{11}$, $\rep3_{12}$, $\rep3_{21}$ and $\rep3_{22}$.

Due to the matter content of our model, the only tensor products that are relevant in this work are those
among the three--dimensional representations $\rep3_{11}$ and $\rep3_{12}$, which are obtained as 
\begin{eqnarray}
\begin{pmatrix}
x_1 \\ x_2 \\ x_3 \\
\end{pmatrix}_{{\bf 3}_{11}} \otimes
\begin{pmatrix}
y_1 \\ y_2 \\ y_3 \\
\end{pmatrix}_{{\bf 3}_{11}} 
=\begin{pmatrix}
x_1y_1 \\ x_2y_2 \\ x_3y_3 \\
\end{pmatrix}_{{\bf 3}_{12}} \oplus
\begin{pmatrix}
x_2y_3+x_3y_2 \\ x_3y_1+x_1y_3 \\ x_1y_2+x_2y_1 \\
\end{pmatrix}_{{\bf 3}_{12}} 
\oplus
\begin{pmatrix}
x_2y_3-x_3y_2 \\ x_3y_1-x_1y_3 \\ x_1y_2-x_2y_1 \\
\end{pmatrix}_{{\bf 3}_{22}} ,
\end{eqnarray}
\begin{eqnarray}
\begin{pmatrix}
x_1 \\ x_2 \\ x_3 \\
\end{pmatrix}_{{\bf 3}_{12}} \otimes
\begin{pmatrix}
y_1 \\ y_2 \\ y_3 \\
\end{pmatrix}_{{\bf 3}_{12}} 
=\begin{pmatrix}
x_1y_1 \\ x_2y_2 \\ x_3y_3 \\
\end{pmatrix}_{{\bf 3}_{11}} \oplus
\begin{pmatrix}
x_2y_3+x_3y_2 \\ x_3y_1+x_1y_3 \\ x_1y_2+x_2y_1 \\
\end{pmatrix}_{{\bf 3}_{11}} 
\oplus
\begin{pmatrix}
x_2y_3-x_3y_2 \\ x_3y_1-x_1y_3 \\ x_1y_2-x_2y_1 \\
\end{pmatrix}_{{\bf 3}_{21}} ,
\end{eqnarray}
and finally 
\begin{eqnarray}
\begin{pmatrix}
x_1 \\ x_2 \\ x_3 \\
\end{pmatrix}_{{\bf 3}_{11}} \otimes
\begin{pmatrix}
y_1 \\ y_2 \\ y_3 \\
\end{pmatrix}_{{\bf 3}_{12}} 
&=&
\begin{pmatrix}
x_1y_1 + x_2y_2 + x_3y_3
\end{pmatrix}_{{\bf 1}_{0}} \oplus
\begin{pmatrix}
x_1y_1+\omega ^2x_2y_2+\omega x_3y_3 \\ \omega x_1y_1+\omega 
^2x_2y_2+x_3y_3
\end{pmatrix}_{{\bf 2}_1} \nonumber \\ 
& \oplus &
\begin{pmatrix}
x_1y_2+\omega ^2x_2y_3+\omega x_3y_1 \\ \omega x_1y_3+\omega 
^2x_2y_1+x_3y_2
\end{pmatrix}_{{\bf 2}_2} \oplus
\begin{pmatrix}
x_1y_3+\omega ^2x_2y_1+\omega x_3y_2 \\ \omega x_1y_2+\omega 
^2x_2y_3+x_3y_1
\end{pmatrix}_{{\bf 2}_3} \nonumber \\ 
& \oplus &
\begin{pmatrix}
x_1y_3+x_2y_1+x_3y_2 \\ x_1y_2+x_2y_3+x_3y_1
\end{pmatrix}_{{\bf 2}_4},
\end{eqnarray}
where $\omega=e^{2\pi\I/3}$. It follows that the only products of $\Delta(54)$ triplets 
up to trilinear order that yield invariant combinations are $\rep3_{11}\otimes\rep3_{12}$,
$\rep3_{11}\otimes\rep3_{11}\otimes\rep3_{11}$ and $\rep3_{12}\otimes\rep3_{12}\otimes\rep3_{12}$.
The latter two products lead to two invariant singlets $\rep1_0$ each.

\end{appendix}



\providecommand{\bysame}{\leavevmode\hbox to3em{\hrulefill}\thinspace}
\frenchspacing
\newcommand{\origttfamily}{}
\let\origttfamily=\ttfamily
\renewcommand{\ttfamily}{\origttfamily \hyphenchar\font=`\-}

\end{document}

%% file: Z3fixedpoints.pdf_t
\begin{picture}(0,0)%
\includegraphics{Z3fixedpoints.pdf}%
\end{picture}%
\setlength{\unitlength}{11188sp}%
\begingroup\makeatletter\ifx\SetFigFont\undefined%
\gdef\SetFigFont#1#2#3#4#5{%
  \reset@font\fontsize{#1}{#2pt}%
  \fontfamily{#3}\fontseries{#4}\fontshape{#5}%
  \selectfont}%
\fi\endgroup%
\begin{picture}(1070,730)(1617,51)
\put(1944,115){\makebox(0,0)[lb]{\smash{{\SetFigFont{11}{13.2}{\rmdefault}{\mddefault}{\updefault}$z_{f,0}$}}}}
\put(2425,130){\makebox(0,0)[lb]{\smash{{\SetFigFont{11}{13.2}{\rmdefault}{\mddefault}{\updefault}$e_1$}}}}
\put(2214,313){\makebox(0,0)[lb]{\smash{{\SetFigFont{11}{13.2}{\rmdefault}{\mddefault}{\updefault}$z_{f,1}$}}}}
\put(1632,580){\makebox(0,0)[lb]{\smash{{\SetFigFont{11}{13.2}{\rmdefault}{\mddefault}{\updefault}$e_2$}}}}
\put(1883,489){\makebox(0,0)[lb]{\smash{{\SetFigFont{11}{13.2}{\rmdefault}{\mddefault}{\updefault}$z_{f,2}$}}}}
\end{picture}%

%% file: Delta54inZ3.pdf_t
\begin{picture}(0,0)%
\includegraphics{Delta54inZ3.pdf}%
\end{picture}%
\setlength{\unitlength}{11188sp}%
\begingroup\makeatletter\ifx\SetFigFont\undefined%
\gdef\SetFigFont#1#2#3#4#5{%
  \reset@font\fontsize{#1}{#2pt}%
  \fontfamily{#3}\fontseries{#4}\fontshape{#5}%
  \selectfont}%
\fi\endgroup%
\begin{picture}(884,730)(1662,32)
\put(1807,532){\makebox(0,0)[lb]{\smash{{\SetFigFont{11}{13.2}{\rmdefault}{\mddefault}{\updefault}$m=2$}}}}
\put(1677, 87){\makebox(0,0)[lb]{\smash{{\SetFigFont{11}{13.2}{\rmdefault}{\mddefault}{\updefault}$m=0$}}}}
\put(2432,615){\makebox(0,0)[lb]{\smash{{\SetFigFont{11}{13.2}{\rmdefault}{\mddefault}{\updefault}$q=1$}}}}
\put(2236,297){\makebox(0,0)[lb]{\smash{{\SetFigFont{11}{13.2}{\rmdefault}{\mddefault}{\updefault}$m=1$}}}}
\end{picture}%